%% file: main.tex
\documentclass[conference]{IEEEtran}
\IEEEoverridecommandlockouts

\usepackage{cite}
\usepackage{amsmath,amssymb,amsfonts}
\usepackage{algorithmic}
\usepackage{graphicx}
\usepackage{textcomp}
\usepackage{xcolor}

\usepackage{outlines}
\usepackage{amsmath}
\usepackage{standalone}
\usepackage{multirow}
\usepackage{adjustbox}
\usepackage{braket}
\usepackage{tikz}
\usepackage{outlines}
\usetikzlibrary{shapes.geometric, arrows.meta, positioning, calc, fit, backgrounds}

\usepackage{pgfplots}
\usepackage{pgfplotstable}
\usetikzlibrary{patterns}
\pgfplotsset{compat=1.18}
\usepackage{xcolor}

\usepackage[letterpaper, top=1in, bottom=1in, left=0.75in, right=0.75in, includeheadfoot=false, headheight=0pt, headsep=0pt, footskip=0pt]{geometry} 
\usepackage{booktabs}
\usepackage{array}
\usepackage{multirow}
\usepackage[table]{xcolor}
\usepackage{tabularx}
\usepackage{amsmath}

\definecolor{lvl5color}{HTML}{D53E4F}
\definecolor{lvl4color}{HTML}{FC8D59}
\definecolor{lvl3color}{HTML}{FEE08B}
\definecolor{lvl2color}{HTML}{E6F598}
\definecolor{lvl1color}{HTML}{3288BD}

\newcolumntype{L}{>{\raggedright\arraybackslash}X}
\renewcommand{\arraystretch}{0.78}
\def\BibTeX{{\rm B\kern-.05em{\sc i\kern-.025em b}\kern-.08em
    T\kern-.1667em\lower.7ex\hbox{E}\kern-.125emX}}
\begin{document}

\title{Architecting Instruction Execution for Scalable Fault-Tolerant Quantum Computing
}

\author{
\IEEEauthorblockN{
Dmitry Filippov\textsuperscript{*},
Justin Hogaboam\textsuperscript{\dag},
Mathias Soeken\textsuperscript{\dag},
Raghu Gatta\textsuperscript{\dag},
Prakash Murali\textsuperscript{*},
Matthias Troyer\textsuperscript{\dag}
}
\IEEEauthorblockA{
\textsuperscript{*}\textit{Department of Computer Science and Technology, University of Cambridge} \quad
\textsuperscript{\dag}\textit{Microsoft}
}
\IEEEauthorblockA{
\{df476, pm830\}@cam.ac.uk \quad \{jhogaboam, Mathias.Soeken, rgatta, matthias.troyer\}@microsoft.com
}
}

\maketitle

\begin{abstract}
Scaling fault-tolerant quantum computing (FTQC) requires orchestrating millions of operations continuously across a tightly coupled classical-quantum control stack. A critical challenge in this orchestration is determining the optimal hardware-software boundary, i.e., the Quantum Instruction Set Architecture (QISA).
We focus on cryogenic, fixed-connectivity qubit platforms, such as superconducting and topological Majorana qubits. For these, existing QISA choices either operate at too low a level, violating cryogenic thermal budgets at scale, or at too high a level, imposing unrealizable complexity on the control engine.
To reason about these tradeoffs, we introduce the ARIES instruction set hierarchy, which categorizes QISAs into distinct abstraction levels. To systematically evaluate these levels, we develop an end-to-end evaluation toolflow featuring a cycle-accurate classical driver and hardware engine simulator. Using eight quantum applications across superconducting and topological qubit platforms, we measure instruction bandwidth, transceiver power dissipation, and measurement processing overheads. We demonstrate that while physical-level abstractions exceed the thermal limits of dilution refrigerators by orders of magnitude, canonical-level abstractions significantly underestimate hardware routing congestion and execution runtimes. Our Level 4 QISA (local logical operations) offers an optimal architectural balance. As quantum hardware platforms scale toward FTQC in the coming years, our findings provide concrete architectural guidance for the design of their classical control stacks and instruction interfaces.
\end{abstract}


\section{Introduction}
Quantum computing is transitioning from the noisy intermediate-scale quantum (NISQ) to \emph{fault-tolerance}. Fault-tolerant quantum computing (FTQC) will rely on quantum error correction (QEC), where collections of physical qubits are entangled to encode resilient \emph{logical} qubits. Logical qubits have now been demonstrated across multiple hardware platforms, including superconducting circuits, trapped ions, and neutral atoms~\cite{Acharya2023, 2404.02280, Bluvstein2024, 2411.11822, 2404.16728}. 
To achieve \emph{practical quantum advantage} over classical computing for commercially or scientifically useful applications, quantum computers must scale to thousands of logical qubits~\cite{resource_estimation_paper, Daley2022, gidney_ekera, 2404.16351}. Applications in chemistry, materials science, and cryptography require executions of billions of logical operations continuously over days or weeks~\cite{resource_estimation_paper}. 


\definecolor{modernSlate}{RGB}{241, 245, 249}
\definecolor{modernIndigo}{RGB}{238, 242, 255}
\definecolor{techIndigo}{RGB}{79, 70, 229}
\definecolor{modernCoral}{RGB}{244, 63, 94}
\definecolor{darkSlate}{RGB}{30, 41, 59}
\begin{figure}[t]
    \centering
    \resizebox{0.92\columnwidth}{!}{%
    \begin{tikzpicture}[
        >=Stealth,
        boxbase/.style={draw=darkSlate, line width=1pt, rounded corners=4pt, align=center, font=\sffamily\small, inner sep=5pt, minimum width=6.5cm},
        appbox/.style={boxbase, fill=modernSlate},
        driverbox/.style={boxbase, fill=modernSlate},
        enginebox/.style={boxbase, fill=modernIndigo!20},
        qpubox/.style={boxbase, fill=modernIndigo!20},
        pathlabel/.style={font=\sffamily\scriptsize, align=center, fill=white, inner sep=2pt, rounded corners=2pt},
        hostinternal/.style={font=\sffamily\scriptsize, align=center, fill=modernSlate, inner sep=2pt, rounded corners=2pt},
        cryointernal/.style={font=\sffamily\scriptsize, align=center, fill=modernIndigo!90, inner sep=2pt, rounded corners=2pt},
        env/.style={rounded corners=6pt, inner sep=8pt, dashed, line width=0.6pt}
    ]
    \node[appbox] (app) {
        \textbf{Quantum Application}\\[-2pt]
        \footnotesize (e.g., Quantum chemistry, Shor's algorithm)
    };
    \node[appbox, below=0.4cm of app] (compiler) {
        \textbf{Compiler}
    };
    \node[driverbox, below=0.7cm of compiler] (driver) {
        \textbf{Classical Software Driver}\\[1pt]
        \footnotesize
        \begin{tabular}{l}
        $\bullet$ JIT instruction translator\\
        $\bullet$ Logical Pauli frame tracking\\
        $\bullet$ Classical control flow
        \end{tabular}
    };
    \node[enginebox, below=2.5cm of driver] (engine) {
        \textbf{Control Engine}\\[-2pt]
        \scriptsize (e.g., FPGA, Cryo-CMOS)\\[1pt]
        \footnotesize
        \begin{tabular}{l}
        $\bullet$ SerDes (Serializer/Deserializer)\\
        $\bullet$ Microcode or SIMD generator\\
        $\bullet$ QEC decoder with physical Pauli frame tracking
        \end{tabular}
    };
    \node[qpubox, below=1.2cm of engine] (qpu) {
        \textbf{Quantum Processor (QPU)}\\[-2pt]
        \footnotesize Array of physical qubits
    };
    \draw[->, thick, draw=darkSlate!40] (app) -- (compiler);
    \draw[->, thick, draw=darkSlate!40] (compiler) -- node[right, hostinternal, xshift=2pt] {
        \textbf{Ahead of time compilation}\\ARIES Level 5: Canonical IR
    } (driver);
    \coordinate (mid_boundary) at ($(driver.south)!0.5!(engine.north)$);
    \draw[line width=1.2pt, dashed, color=modernCoral]
        ([xshift=-4.2cm]mid_boundary) -- ([xshift=4.2cm]mid_boundary);
    \node[above=3pt, align=center, font=\sffamily\footnotesize, fill=white, inner sep=2pt, rounded corners=2pt] at (mid_boundary) {
        \bfseries\color{modernCoral!80!black}SOFTWARE-HARDWARE BOUNDARY\\
        \bfseries\color{modernCoral!80!black}(QISA Design Space)
    };
    \node[below=3pt, align=center, font=\sffamily\scriptsize, fill=white, inner sep=2pt, rounded corners=2pt] at (mid_boundary) {
        \color{darkSlate}\textbf{Bottleneck:} Interconnect Capacity ($BW_{cap}$),\\
        \color{darkSlate}Thermal Limit ($P_{budget}$) \&\\
        \color{darkSlate}Driver's Measurement Processing Overhead
    };
    \draw[->, line width=1.5pt, draw=techIndigo]
        ([xshift=-2.5cm]driver.south) -- node[left=3pt, pathlabel, align=right] {
            \textbf{ISSUE PATH}\\QISA Instructions\\
            \textit{(ARIES Levels 1--5)}\\
            \color{techIndigo}$\mathbf{BW_{issue}}$
        } ([xshift=-2.5cm]engine.north);
    \draw[<-, line width=1.5pt, draw=modernCoral]
        ([xshift=2.5cm]driver.south) -- node[right=3pt, pathlabel, align=left] {
            \textbf{RETURN PATH}\\Algorithmic Meas. \&\\
            Decoded Syndromes\\
            \color{modernCoral}$\mathbf{BW_{return}^{algo}}$
        } ([xshift=2.5cm]engine.north);
    \draw[->, line width=1.5pt, draw=techIndigo]
        ([xshift=-2cm]engine.south) -- node[left=3pt, cryointernal, align=right] {
            \textbf{PHYSICAL OPS}\\(ARIES Level 1)
        } ([xshift=-2cm]qpu.north);
    \draw[<-, line width=1.5pt, draw=modernCoral]
        ([xshift=2cm]engine.south) -- node[right=3pt, cryointernal, align=left] {
            \textbf{FAST QEC LOOP}\\Raw Parity Readouts\\
            \color{modernCoral}$\mathbf{BW_{return}^{QEC}}$
        } ([xshift=2cm]qpu.north);
    \begin{scope}[on background layer]
        \node[env, draw=darkSlate!20, fill=modernSlate, fit=(app)(compiler)(driver), inner xsep=1cm] (host_env) {};
        \node[env, draw=darkSlate!10, fill=modernIndigo!90, fit=(engine)(qpu), inner xsep=1cm] (device_env) {};
    \end{scope}
    \end{tikzpicture}%
    }
    \caption{FTQC instruction execution flow. 
    }
    \label{fig:driver_engine}
\end{figure}
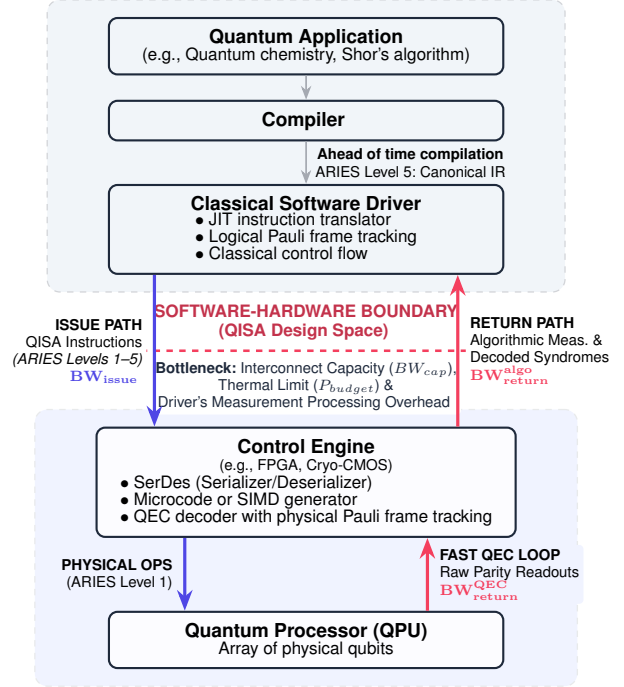

Executing FTQC applications requires continuous coordination between a classical software \emph{driver} and a hardware control \emph{engine}. Figure~\ref{fig:driver_engine} illustrates these components. A program written in a high-level language such as Q\# is compiled into low level operations which the driver issues to the engine at runtime. The driver has to perform translations to support adaptive control flow, decompose remote operations into local sequences and apply Pauli frame corrections~\cite{littinski_game}.
Full pre-compilation of practical-scale quantum applications leads to prohibitive binary sizes~\cite{JAVADIABHARI20152}; therefore, just-in-time compilation is used to stream batches of instructions into the quantum device, much like a classical co-processor~\cite{Britt2017-is, Wong2025-ef}.
The engine executes instructions on the quantum device and extracts QEC syndrome measurements (parity checks). These results are streamed back to the driver for error decoding and control flow resolution. Issue and return paths must operate continuously at the speed of the quantum hardware to avoid stalling QEC cycles or application logic.



Quantum Instruction Set Architecture (QISA) choices dictate key tradeoffs across the FTQC stack. A low-level QISA that exposes physical operations as in current QEC demonstrations~\cite{Acharya2023, eth_surface_code} requires the driver to issue instructions for all qubits every QEC cycle, even for idling qubits. For cryogenic platforms, this bandwidth translates to thermal power dissipation at the driver-engine interface that may violate cooling power limits of dilution fridges. 
A high-level logical QISA, on the other hand, compresses bandwidth but may push impractical complexity onto the engine. Designing a scalable control stack therefore requires a systematic analysis of the QISA design space across abstraction levels, platforms, and application workloads.

\textbf{Scope:} We focus on cryogenic platforms with fixed two-dimensional qubit connectivity, namely superconducting and topological Majorana qubits, where the driver-engine interface must physically cross a thermal boundary into a dilution refrigerator. We analyze the impact of QISA choice on this interface's bandwidth and cooling budget, as well as on engine hardware complexity and driver runtime overheads. Extending this analysis to platforms with different constraints, such as trapped ions or neutral atoms, is valuable future work.


\begin{figure}[t]
    \centering
    \resizebox{0.9\columnwidth}{!}{%
    \begin{tikzpicture}[
        >=Stealth,
        boxbase/.style={draw=darkSlate, line width=1pt, rounded corners=4pt, align=center, font=\sffamily\small, inner sep=5pt, minimum width=7.2cm},
        level54/.style={boxbase, fill=modernSlate},
        level3/.style={boxbase, fill=modernSlate!80!modernIndigo!50},
        level21/.style={boxbase, fill=modernIndigo!20},
    ]
    \node[level54] (l5) {
        \textbf{Level 5: Canonical Operations}\\
        \footnotesize Architecture-agnostic algorithmic instructions\\[-1pt]
        \scriptsize\textit{QISA choice in~\cite{littinski_game, resource_estimation_paper}}
    };
    \node[level54, below=0.35cm of l5] (l4) {
        \textbf{Level 4: Local Logical Operations}\\
        \footnotesize Composite logical gates which act on local logical qubits\\[-1pt]
        \scriptsize\textit{\textbf{\color{techIndigo}$\bigstar$ Our work's recommended QISA level}}
    };
    \node[level3, below=0.35cm of l4] (l3) {
        \textbf{Level 3: QEC-Native Logical Operations}\\
        \footnotesize Native gates and measurements of the QEC scheme\\[-1pt]
        \scriptsize\textit{QISA choice in~\cite{edge_disjoint_paths, fowler_gidney_2018, tannu_micro}}
    };
    \node[level21, below=0.35cm of l3] (l2) {
        \textbf{Level 2: QEC Micro-Operations}\\
        \footnotesize Stabilizer patch building blocks\\[-1pt]
        \scriptsize\textit{QISA choice in~\cite{xqsim}}
    };
    \node[level21, below=0.35cm of l2] (l1) {
        \textbf{Level 1: Physical Operations}\\
        \footnotesize Individual or SIMD physical gates and measurements\\[-1pt]
        \scriptsize\textit{QISA choice in~\cite{Acharya2023, eth_surface_code}}
    };
    \draw[->, thick, draw=darkSlate!60] (l5) -- (l4);
    \draw[->, thick, draw=darkSlate!60] (l4) -- (l3);
    \draw[->, thick, draw=darkSlate!60] (l3) -- (l2);
    \draw[->, thick, draw=darkSlate!60] (l2) -- (l1);
    \end{tikzpicture}%
    } 
    \caption{Our ARIES QISA hierarchy with five abstraction levels from hardware-agnostic canonical operations down to physical SIMD instructions. Our work quantifies design tradeoffs at each level and proposes Level 4 ISAs to balance software and control overheads.}
    \label{fig:aries_hierarchy}
\end{figure}
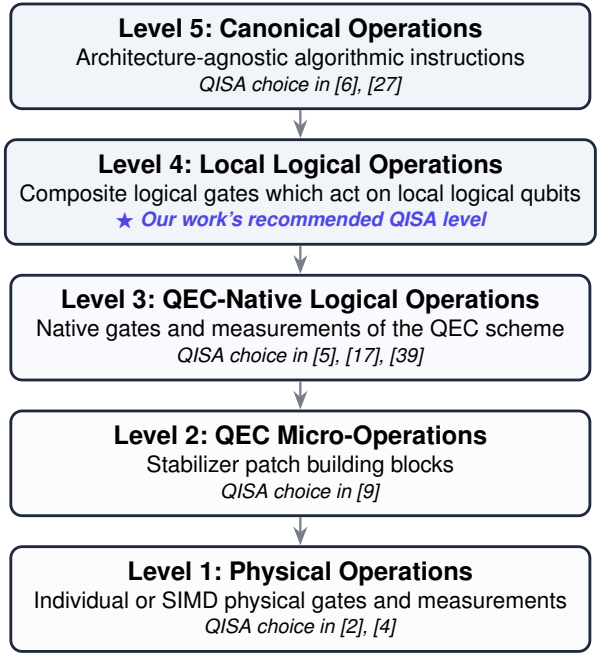

To address these challenges, we develop the ARIES instruction set hierarchy. Figure~\ref{fig:aries_hierarchy} lays out the taxonomy of operations capturing five abstraction levels, from canonical architecture-agnostic operations at the top, down to physical operations at the bottom. 
To evaluate design tradeoffs across these levels, we develop the toolflow shown in Figure \ref{fig:toolflow}. It takes as input a quantum application and an architecture configuration (qubit and QEC characteristics, QISA level)
and produces end-to-end metrics including estimated application execution cycles, interconnect bandwidth and power. We evaluate QISA design choices across eight practically-relevant quantum applications and two qubit platforms (superconducting qubits and topological qubits). Our contributions are:
\begin{itemize}
\item On cryogenic platforms like superconducting or topological qubits, physical QISA choices are acceptable for current demonstrations~\cite{Acharya2023, eth_surface_code}. But, for practical applications, physical QISAs violate cooling power constraints of dilution fridges by over $4000\times$. Up-leveling the QISA is a strict requirement for scalable quantum computation. 
\item Prior work~\cite{tannu_micro} identifies issue bandwidth as a bottleneck and proposes a Level 3 ISA, focusing exclusively on the issue path. We show that the return path is also highly sensitive to the QISA level. Level 4 QISAs offer 2.6$\times$ (up to 5.2$\times$) return bandwidth reductions on top of Level 3 while also speeding up quantum chemistry and factoring kernels by 1.2-1.4$\times$.
\item  Prior works~\cite{littinski_game, resource_estimation_paper} use highly abstract CISC-like operations (at Level 5) and assume variable bitwidth instructions and dynamic qubit management and routing in hardware. Level 4 ISAs provide an optimal architectural balance, allowing software to handle such functions more flexibly and supporting simple, deterministic RISC-like instructions. 
\end{itemize}


\section{Background}
\textbf{Qubit Technologies:} 
Superconducting qubits are typically implemented as transmons~\cite{superconducting_qubits}. They feature fast nanosecond-scale gates and readout but are constrained to fixed two-dimensional physical qubit layouts. 
 The control engine resides at 4K inside a dilution refrigerator, imposing a strict cooling budget on the classical-quantum interface~\cite{tannu_micro, superconducting_qubits, krinner}. 
 Topological qubits, based on Majorana fermions~\cite{majorana_roadmap, resource_estimation_paper}, are expected to achieve lower physical error rates than existing technologies while supporting fast nanosecond-scale operation times. Like superconducting systems, topological architectures require cryogenic control with similar thermal constraints. 

\textbf{Quantum Error Correction:} In FTQC, physical qubits are grouped into \emph{logical qubits} using a QEC code. The \emph{code distance} $d$ controls the number of physical qubits and exponentially suppresses the logical error rate, provided physical qubits operate below a threshold error rate~\cite{fowler_gidney_2018}. Error syndromes are extracted periodically via parity measurements; one extraction round completes in a platform-dependent time~$t_r$, and $d$ consecutive rounds establish a logical cycle time $t_c = d \times t_r$. Practical applications require $d \approx 15$--$25$~\cite{resource_estimation_paper}, implying that a logical cycle takes several microseconds.

The surface code~\cite{fowler_gidney_2018} is a leading QEC choice for both superconducting~\cite{Acharya2023} and topological Majorana platforms~\cite{Chao2020}. Hastings-Haah code~\cite{Hastings2021, Haah2022} is an alternative QEC choice for Majorana platforms that uses fewer syndrome extraction rounds per logical cycle ($t_r = 3 \times t_\text{meas}$ vs.\ $20 \times t_\text{meas}$ for surface codes) at the cost of more physical qubits per logical qubit. Logical operations on these codes are implemented via \emph{lattice surgery}: ancilla qubits measure joint Pauli parities between neighboring logical patches (e.g., MeasZZ, MeasXX). Each such measurement requires $O(d)$ syndrome extraction rounds and its result must be processed to apply corrections. ARIES Levels 2 and 3 operate at this level, making it specific to a particular QEC scheme. Levels 4 and 5 are on the other hand QEC-agnostic. 

\textbf{Pauli Frame Tracking:} In QEC, parity check measurements allow us to determine the location and type of errors on a set of physical qubits. With surface codes, these errors don't have to be corrected physically. They can be tracked in software and used to correctly interpret future measurement outcomes. This is referred to as Pauli frame tracking, where a per-qubit record of the accumulated X, Z or Y error is maintained and updated as computation progresses~\cite{Chamberland2018faulttolerant, Riesebos2017PauliFF}. 

\textbf{Clifford vs. non-Clifford operations:} Clifford operations (e.g., CNOT, Hadamard, S, X, Y, Z) are those that are natively supported by QEC schemes. Non-Clifford operations (e.g., T gate or Toffoli gate) are required for universality; to implement these operations fault-tolerantly, state injection is used, which requires significantly more compute resources and creates synchronization points at which pending Pauli frame corrections must be applied~\cite{windowed_decoding}.


\section{QISA Design Tradeoffs}
\begin{figure}[t]
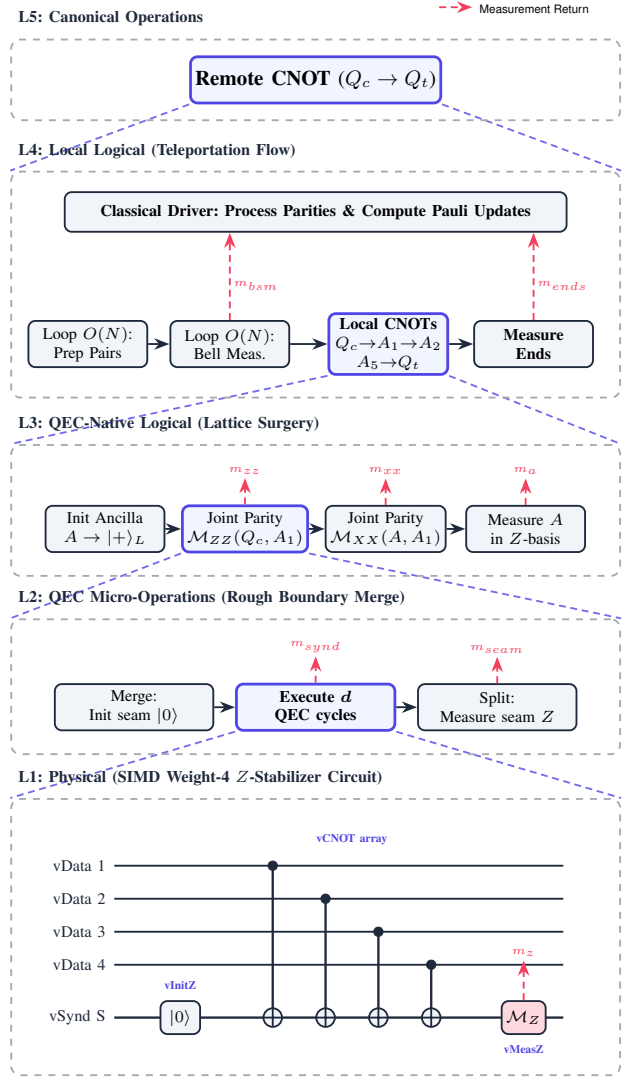

    \centering
    \includestandalone[width=0.93\columnwidth]{images/cnot_heir}
    \caption{ARIES five-level QISA hierarchy illustrated through the decomposition of a remote logical CNOT into physical operations. The protocol assumes edge-disjoint paths layout~\cite{edge_disjoint_paths} at the logical level with a surface code QEC. Red arrows denote measurements that need to be processed either by the driver or engine, depending on the QISA level.}
    \label{fig:cnot_hierarchy}
\end{figure}
\begin{figure}[t]
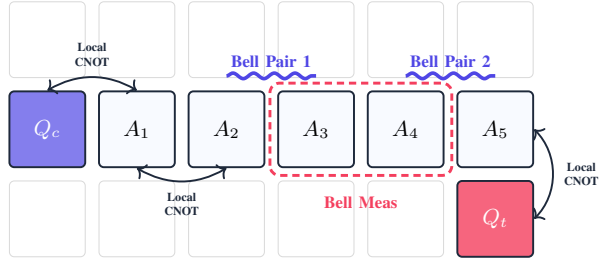

    \centering
    \includestandalone[width=0.93\columnwidth]{images/tiled_surface_code}
    \caption{A routing path for a remote CNOT between $Q_c$ and $Q_t$ on a $3 \times 6$ surface code tile array using the protocol in \cite{edge_disjoint_paths}. Ancilla qubits $A_1$--$A_5$ form the routing path, with Bell pairs prepared between non-adjacent pairs and entanglement swaps (Bell measurements) on adjacent pairs to extend entanglement across the path.}
    \label{fig:tiled_surface_code}
\end{figure}
\subsection{Instruction Issue and Measurement Processing}
During algorithm execution, operations must be continuously executed at the logical clock frequency of the QEC cycle, without stalling the quantum processor. 
A stall not only causes performance penalty, but also degrades the quality of computation by missing the opportunity to detect and correct errors that happen on the qubits during the stall. Therefore, asynchronous instruction issue is required. The driver dispatches instructions to the engine and continues preparing subsequent instructions without waiting for the previous instructions to complete. However, measurement results returning from the engine carry Pauli frame updates that affect future instructions. The Pauli frame corrections can be commuted through the Clifford gates and can be deferred, allowing us to issue Clifford gates without waiting for the measurement results. However, for non-Clifford operations, all pending Pauli frame corrections must be applied before issue, creating a synchronization point in the pipeline~\cite{divincenzo_aliferis_2007, windowed_decoding, Chamberland2018faulttolerant}. 
The choice of QISA abstraction dictates how much measurement processing the driver must complete at each such synchronization point. 

Measurements from the qubits serve multiple purposes: algorithmic measurements, logical qubit measurements, and QEC measurements. Algorithmic measurements are those that are present in the user program (or inserted by the compiler) and are handled by the program's logic. Logical qubit measurements are those that could arise from decompositions of algorithmic operations in terms of QISA operations. For example, Figure \ref{fig:cnot_hierarchy}, shows the decomposition of an algorithmic CNOT gate into a sequence of native measurement operations which can be run on the surface code. In this case, the measurement results are used to steer the logical qubits to the correct states. QEC measurements directly arise from syndrome measurement sequences used in QEC and result in Pauli frame updates. The QISA level determines which of these measurement categories the driver must process. With low-level ISAs, the driver must process all measurements; with a very high-level ISA, the driver needs to only handle algorithmic measurements. 

For each measurement that needs to be processed in the driver, we require a future to hold the result of the asynchronous in-flight instruction. A very low-level ISA abstraction could overwhelm the driver by demanding millions of futures and their associated management overhead, potentially slowing down the application and instruction issue cycle. If measurement processing at the driver cannot keep pace with the logical clock frequency, instruction stalls and decoding backlog can accumulate. Such backlogs can lead to exponential slowdown in the computation, as unresolved syndromes compound across cycles~\cite{RevModPhys.87.307}. Abstracting the QISA allows us to reduce the volume of measurement processing in software. 


\subsection{Interface Bandwidth}
The driver-engine interface must sustain sufficient bandwidth in both directions to keep up with the quantum processor. On the issue path, the driver issues QISA instructions to the engine at a rate determined by the number of active logical qubits and the logical clock frequency. On the return path, the driver must process measurement results from the engine. A physical-level QISA requires the maximum bandwidth since the driver must issue instructions to every physical qubit in every QEC cycle and handle the measurement results. 
Abstracting the QISA to the logical level reduces this significantly by introducing \emph{implicit idling}~\cite{tannu_micro}, which means that the engine automatically issues QEC instructions for every logical qubit and the driver only handles algorithmic instructions. QEC syndrome extraction circuits often repeat the same structured pattern of gates across all data qubits every cycle, making them easy to parallelize through SIMD-style control.



Depending on the QISA abstraction, the types and volume of measurements returned to the driver vary, from every raw syndrome bit at the 
physical level to only application-visible results at higher abstractions. Both issue and 
return bandwidths must be sustained over the physical interconnect between the driver and 
engine, making the choice of QISA abstraction a key determinant of whether practical 
interconnect implementations are feasible.
\subsection{Transceiver Power}
On superconducting or topological qubits, the driver-engine interface connects room temperature electronics to cryogenic hardware that resides in a dilution fridge. Qubits are typically housed in the milli-Kelvin stage of this fridge. A serializer/deserializer (SerDes) is required at this boundary to convert parallel instruction and measurement data into a serial stream. This serial stream is then transmitted over a limited number of cabling lines between temperature stages. Without SerDes, a massive number of control lines, at least one per physical qubit, would be required to connect into the fridge. Since each cable conducts heat into the cryostat, the number of lines is constrained by thermal conductivity. 

The power consumed by the SerDes at the cryogenic interface acts as a lower bound on the total engine power. This power scales with the sustained bandwidth at the driver-engine interface, $P_{\text{SerDes}} = BW \times E_{\text{bit}}$, where $E_{\text{bit}}$ is the energy per bit of the transceiver. For cryogenic stages cooled to 4K, the available cooling power is strictly limited (approximately 1.5W for current dilution refrigerators~\cite{krinner}). QISAs which demand high interface bandwidth require high SerDes power and can exceed the cooling budget of the dilution fridge.

\subsection{Control Footprint per Instruction}
The \emph{control footprint} of a QISA instruction is the set of engine resources (micro and physical operation count, execution cycles, and hardware state) required to execute a single instruction.
While bandwidth considerations favour higher-level instruction sets, control footprint increases as we increase abstraction, creating a fundamental tension between the two. Consider a remote CNOT operation. If such an operation is used as the QISA, then the driver can issue a single instruction specifying any two logical qubits in the system. The engine must orchestrate the operation by allocating ancilla qubits along the path, preparing Bell pairs, doing Bell measurements and applying corrections based on the measurement results. This results in a variable number of lower-level operations, each producing asynchronous measurement values that must be processed internally. 
The micro-operation count, cycle count, and ancilla requirements all depend on the distance between the operands, making the control footprint unbounded and operand-dependent.
Similarly, canonical instruction sets that support arbitrary multi-qubit Pauli measurements (as commonly assumed in resource estimation studies~\cite{resource_estimation_paper}) require instruction widths proportional to the number of logical qubits. These studies typically assume such operations complete in a single logical cycle, but their physical implementation requires decomposition into a variable-length sequence of local operations. 

On the other hand, if an instruction set only allows fixed-width, local operations, it can offer a deterministic control footprint. Each instruction acts on a fixed number of neighboring logical qubits, generates a bounded number of micro-operations and completes in a fixed number of cycles.
This enables fixed-size hardware decoders, predictable pipelining in control hardware and simplified hardware provisioning.
Variable cost tasks such as ancilla allocation, routing path selection etc. can be handled by the driver in software, with more computational flexibility. 
The analogy in classical computing is a fixed-width 32-bit add instruction versus a variable-width $n$-bit add where $n$ varies up to an application-dependent length. The former enables efficient, predictable hardware while the latter demands unbounded datapath resources.

\begin{table*}[ht]
\centering
\caption{ARIES Instruction Set Hierarchy: Operations at Each ISA Level.}
\label{tab:aries_isa}
\renewcommand{\arraystretch}{1.4}
\begin{tabularx}{\textwidth}{>{\bfseries}p{3.6cm} L}
\toprule
\textbf{Level} & \textbf{Operations} \\
\midrule

Level 5: Canonical Operations &
\texttt{Allocate} $Q$ /
\texttt{Free} $Q$;\quad
$R_x(\theta)$ / $R_y(\theta)$ / $R_z(\theta)$;\quad
\texttt{CanonicalClifford} $Q, \mathrm{Op}$;\quad
\texttt{MeasPauli} $Q_0, \ldots, Q_{n-1}, P$;\quad
\texttt{RemoteControlledClifford} $Q_c, Q_t, \mathrm{Op}$ \\

\midrule

Level 4: Local Logical Operations &
\texttt{Init} $Q, P$ /
\texttt{Meas} $Q, P$;\quad
\texttt{InitBell} $Q_0, Q_1$ /
\texttt{MeasureBell} $Q_0, Q_1$;\quad
\texttt{LocalClifford} $Q, \mathrm{Op}$;\quad
\texttt{LocalCNOT} $Q_0, Q_1$;\quad
\texttt{LocalSWAP} $Q_0, Q_1$;\quad
\texttt{LocalMove} $Q_0, Q_1$;\quad
\texttt{CheckMagic} $L$;\quad
\texttt{ConsumeMagic} $Q_{\mathrm{target}}, L$ \\

\midrule

Level 3: QEC-Native Logical Operations &
\texttt{InitX} $Q$ /
\texttt{InitZ} $Q$;\quad
\texttt{MeasX} $Q$ /
\texttt{MeasZ} $Q$;\quad
\texttt{NativeClifford} $Q, \mathrm{Op}$;\quad
\texttt{MeasXX} $Q_0, Q_1$ /
\texttt{MeasZZ} $Q_0, Q_1$ /
\texttt{MeasXZ} $Q_0, Q_1$;\quad
\texttt{CheckMagic} $L$ \\

\midrule

Level 2: QEC Micro-operations &
\texttt{InitData} $G, P$ /
\texttt{MeasData} $G, P$;\quad
\texttt{ExtractSyndrome} $G$;\quad
\texttt{MergeBoundary} $G_0, G_1, \mathrm{boundary}$ /
\texttt{SplitBoundary} $G, \mathrm{boundary}$;\quad
\texttt{DistillCycle} $F, \mathrm{protocol}$;\quad
\texttt{NotifyMagic} $\mathrm{idx}$ \\

\midrule

Level 1 (All Platforms) &
\texttt{vMeasZ} $q$ /
\texttt{vMeasX} $q$ /
\texttt{vMeasY} $q$;\quad
\texttt{vReset} $q$;\quad
\texttt{vWait} $q, \mathrm{cycles}$;\quad
\texttt{vH} $q$ /
\texttt{vS} $q$ \\

Level 1 (Superconducting) &
\texttt{vRx}$(\theta)$ $q$ /
\texttt{vRy}$(\theta)$ $q$;\quad
\texttt{vVirtualZ}$(\theta)$ $q$;\quad
\texttt{vCZ} $q_{\mathrm{ctrl}}, q_{\mathrm{targ}}$ /
\texttt{vCNOT} $q_{\mathrm{ctrl}}, q_{\mathrm{targ}}$ \\

Level 1 (Topological Majorana) &
\texttt{vJointParity} $q_{\mathrm{list}}, \mathrm{basis}$;\quad
\texttt{vTuneBarrier} $q, V_{\mathrm{gate}}$ \\

\bottomrule
\end{tabularx}
\end{table*}

\section{ARIES Instruction Set Hierarchy}
\label{sec:levels} 
Table \ref{tab:aries_isa} details our ISA hierarchy. It includes five levels to capture different abstraction levels in the stack, ranging from canonical algorithmic operations at the top to physical operations at the bottom. Each level can be expressed in terms of operations supported by its immediate lower level using known gate identities.
For brevity, we omit classical control instructions such as barriers in this paper.

\subsection{Level 5: Canonical Operations}
Canonical operations refer to hardware-agnostic algorithmic instructions used by high-level quantum programming languages and resource estimators~\cite{resource_estimation_paper, littinski_game}. At this level, the quantum computer is viewed as an idealized machine with all-to-all connectivity and continuous-angle gates, completely hiding the qubit layout, scheduling and QEC complexities of the underlying physical system.

The defining architectural characteristic of Level 5 is its \emph{variable control footprint}. For example, a single \texttt{RemoteControlledClifford} (with the Clifford being a common gate such as CNOT) instruction between distant qubits forces the engine to dynamically allocate ancillas, sequence routing operations, and resolve intermediate measurements over many cycles. Quantum resource estimation studies often model these global operations as completing in a single logical cycle~\cite{resource_estimation_paper}, abstracting away the massive routing congestion and latency. A scalable QISA must systematically unpack this abstraction into localized operations to maintain realistic hardware timing. 

\subsection{Level 4: Local Logical Operations}
At Level 4, the abstract canonical operations from Level 5 are mapped to the actual architectural topology of logical qubits on the device. Logical qubit layout is exposed to software to enable it to correctly place and schedule operations. However, this level remains agnostic to the underlying QEC scheme. Because this level restricts execution to \emph{local} interactions between nearest-neighbor logical qubits, the overhead of synthesizing remote operations is shifted to the classical software driver. The driver can orchestrate these using fixed, bounded-footprint operations, ensuring the hardware engine maintains a predictable execution latency.

The defining feature of Level 4 is that it only includes local one and two qubit instructions with deterministic latency. Under these constraints, Level 4 tries to minimize the number of measurement results the driver has to receive and the number of correction gates it has to issue by including instructions such as \texttt{LocalCNOT}, \texttt{ConsumeMagic} and \texttt{InitBell} that would be further decomposed into measurements and corrections on Level 3. 

Level 4's QEC-agnosticism is a valuable architectural property: since the field has not yet converged on a single fault-tolerant QEC scheme, a Level 4 QISA lets the same driver, compiler, and application software target surface codes, Hastings-Haah codes, or future QEC schemes without modification, isolating this still-evolving choice from the engine-facing interface. This decoupling also standardizes the control engine implementation across QEC schemes and even qubit technologies, since Level 4 operations only assume local qubit connectivity rather than any QEC-specific primitive. To our knowledge, no prior QISA proposal is deliberately restricted to this level; we compare against the nearest analogues in Section~\ref{sec:related}.

\subsection{Level 3: QEC-Native Logical Operations}
Level 3 drops the QEC-agnostic abstraction and exposes the \emph{native} fault-tolerant primitives of the QEC scheme. We use 2D surface code~\cite{fowler_gidney_2018} and Hastings-Haah codes ~\cite{Hastings2021, Haah2022} as the examples in this paper, but this level can be extended to other schemes. 

An explicit \texttt{Idle} instruction (perform syndrome extraction on idle logical qubits) is absent from this level. By elevating the QISA to Level 3 or higher, the engine automatically handles background syndrome extraction for any logical patch not actively addressed. This \emph{implicit idling} strictly decouples the classical issue bandwidth from the rapid physical QEC clock rate ($t_r$). 

Prior control-stack proposals situate their QISA at this level: \cite{tannu_micro} abstracts issue-bandwidth for defect-based surface codes on superconducting qubits, while \cite{edge_disjoint_paths} exposes lattice-surgery primitives routed over edge-disjoint paths. Both choices are QEC-specific and, as Section~\ref{sec:bandwidth_power} shows, still leave substantial return-path bandwidth on the table relative to Level 4. 

\subsection{Level 2: QEC Micro-operations}
Micro-operations represent the building blocks of quantum error correction. Micro-ops address physical qubit patches, denoted by patch graphs $G$. The exact structure of the qubit patch (comprising hundreds of physical qubit indices) is not transmitted within each instruction payload. Instead, $G$ is an identifier that is assigned when the logical patch is created. The hardware engine is responsible for internally resolving this identifier into the underlying hardware-specific Level 1 operations bounded to that spatial zone. Level 2 operations must include explicit QEC handling using operations such as \texttt{ExtractSyndrome}. XQSim~\cite{xqsim} models its QISA largely at this level, mixing QEC micro-operations with a small set of QEC-agnostic canonical (Level 5) operations.

\subsection{Level 1: Physical Qubit Operations}
Physical instructions represent the lowest level of the instruction hierarchy. 
While current NISQ execution models rely on individual qubit addressing, we introduce \emph{SIMD physical instructions} to express the massive parallelism inherent to QEC workloads. 
Each instruction operates on one or two qubits, enabling aggressive fan-out of control signals, dramatically reducing interconnect bandwidth. Because underlying platforms rely on vastly different physical mechanisms, the Level 1 QISA is inherently platform-specific. Current experimental demonstrations of the surface code on superconducting qubits~\cite{Acharya2023, eth_surface_code} program hardware directly at this level, composing circuits from individual physical gates and measurements much as in NISQ systems.

\section{ARIES Evaluation Framework}
\label{sec:framework}
\begin{figure}[t]
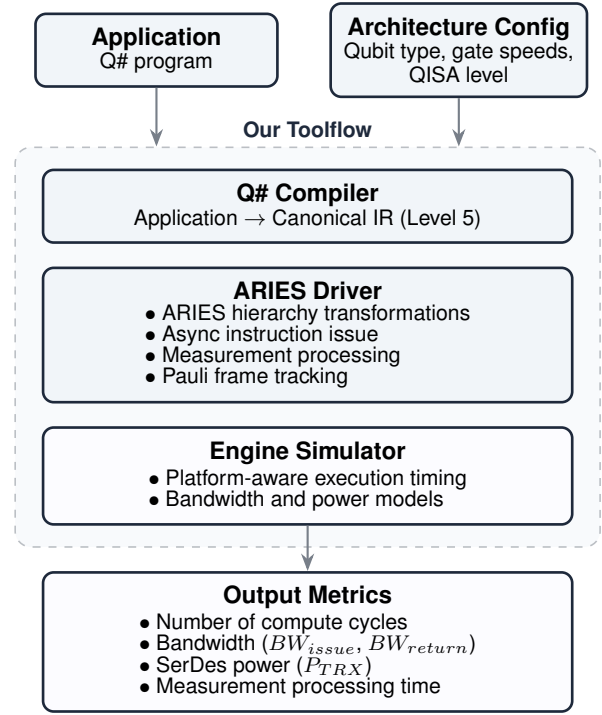

    \centering
    \includestandalone{images/tool_diagram}
    \caption{Our evaluation toolflow. A Q\# application and architecture configuration are fed into the ARIES driver, which transforms instructions through the hierarchy and issues them asynchronously to the engine simulator. Output metrics include number of compute cycles, bandwidth, SerDes power, and measurement processing overhead.}
    \label{fig:toolflow}
\end{figure}

Figure \ref{fig:toolflow} illustrates our evaluation framework. Our toolflow takes Q\# applications as input, along with an architecture configuration that specifies physical qubit parameters, QEC scheme, and the target QISA level. The application is compiled into a canonical Level 5 representation using the Q\# compiler \cite{qsharp}. To ingest this compiled program, we built our own \textbf{cycle-accurate driver and engine simulator, implemented in-house in Rust}. The driver implements a set of transformers that progressively lower instructions from Level 5 to the target QISA level, issues them asynchronously to the engine simulator and processes measurement results, computing necessary corrections. The engine returns measurement values to the driver while tracking instruction counts, allowing us to compute driver-engine bandwidth and transceiver power requirements.

\subsection{Just-In-Time Compilation}
\label{sec:jit}

Our toolflow issues instructions just-in-time (JIT) rather than executing a fully pre-compiled binary, for two reasons. First, FTQC applications have little \emph{algorithmic} control flow, but QEC itself is dynamic: every non-Clifford operation needs a measurement-conditioned correction, magic-state distillation succeeds probabilistically and may need retrying, and interpreting future measurements depends on the Pauli frame accumulated from prior corrections. A static schedule must therefore pre-expand every outcome (intractable, since branches grow exponentially with depth), pad every uncertain operation to its worst case, or resynchronize with the driver at each branch -- reintroducing the round trips pre-compilation was meant to avoid. JIT issue resolves this dynamism only once the relevant measurement result is available.

Second, the instruction streams for larger applications when fully serialized are large, far exceeding cryostat-adjacent storage capacity. For example, the complete Level 4 schedule for factoring is 260 GB, and for nitrogenase ground state estimation it is 400 GB.

\subsection{Driver and Engine Simulation}
\label{sec:modeling}
Our driver assumes a general-purpose classical CPU controls the quantum processor as a co-processor. It traces through the quantum program, progressively lowering instructions through the ARIES hierarchy and issuing them to the engine, applying the gate transformations needed to decompose Level 5 operations to the selected QISA level using known gate identities. To handle asynchronous issue, the driver creates a \texttt{future} for each issued measurement instruction; correction gates dependent on that measurement's outcome are deferred by commuting them through issued Clifford gates, and combined at synchronization points (non-Clifford gates or measurements), minimizing redundant correction instructions.


To complete the evaluation loop, we simulate the execution of a quantum engine. Our simulator receives the instruction stream from the driver, bundles compatible instructions into layers of instructions to be executed in parallel, and executes one layer per clock cycle. The clock is fixed and independent of the driver, so driver latency directly increases layer count through failing to issue all compatible operations before a given layer is executed. The simulator records per-cycle instruction and measurement statistics and returns random measurement outcomes.

\subsection{Bandwidth and Power Model}
\label{sec:bandwidth_power_model}
We model the QISA boundary as a digital interface whose total bandwidth sums the issue path (driver to engine) and the return path (engine to driver): $BW_{total} = BW_{issue} + BW_{return}$.

\textbf{Instruction Issue Bandwidth:} $BW_{issue} = k_L \times s_L \times f_L/\eta$, the product of instructions issued per cycle ($k_L$, profiled empirically from the driver), instruction bit-width ($s_L$, fixed per QISA level), and issue frequency ($f_L$). Crossing physical system boundaries (e.g., room-temperature host to cryogenic engine) requires serializing, transmitting, and deserializing instruction data via a SerDes transceiver; to recover the clock signal and maintain integrity, high-speed SerDes protocols (e.g., Infiniband) use \emph{line encoding} (8b/10b or 64b/66b), injecting redundant non-payload bits. We capture this via line-encoding efficiency $\eta$ (e.g., $\eta \approx 0.8$ for 8b/10b), which reduces effective usable bandwidth.

\textbf{Measurement Return Bandwidth:} Symmetrically, $BW_{return} = m_L \times r_L \times f_L/\eta$, where $m_L$ is the average measurement results returned per cycle (profiled from the driver), $r_L$ is the result bit-width, and $f_L$ is the return frequency (identical to issue frequency).

\textbf{Transceiver Power:} $P_{TRX} = BW_{total} \times E_{bit}$, where $E_{bit}$ is transceiver energy-per-bit. A configuration is viable only if $BW_{total} \le BW_{cap}$ (interconnect capacity) and $P_{TRX} \le P_{budget}$ (platform cooling limit).

\subsection{FPGA Resource Modeling (Microcode)}
The Level 3, 4, and 5 control engines were implemented in System Verilog and synthesized to AMD/Xilinx 7-series FPGA primitives to estimate resource requirements, and we derived an analytical resource model that validates the synthesis results. At 64 logical qubits, engine cost rises monotonically with abstraction level.
\begin{table}[t]
\centering
\caption{LUT resources per QISA level}
\label{tab:LUTs}
\footnotesize
\begin{tabular}{lrcc}
\hline
\textbf{Logical Qubits} & \textbf{Level 3} & \textbf{Level 4} & \textbf{Level 5} \\
\hline
64     &              983 &              2,920  &              5,639   \\
4,096  &              --- &              40,000 &              260,000 \\
\hline
\end{tabular}
\end{table}

From Table \ref{tab:LUTs}, Level 4 (with per-qubit Pauli frame and patch table) costs 40k LUTs at 4,096 logical qubits, while Level 5 must additionally provision an O(N) combinational routing/allocation data path, reaching 260k LUTs -- its router alone needs 217,770 LUTs, exceeding a high-end Kintex-7 FPGA ($\sim$204k LUTs), with per-instruction latency varying nearly 10$\times$ with operand placement. At Level 4, the extra hardware over Level 3 offers QEC abstraction, 1.5-5$\times$ bandwidth improvements (Section \ref{sec:bandwidth_power}), and operand-independent latency for every opcode.

Translating post-synthesis results into a cryo-CMOS ASIP ($\sim$10 NAND2 gate-equivalents/LUT) for realistic cryogenic deployment, the 4,096-qubit Level 4 engine is $\sim$0.5M gate-equivalents, dissipating 10-20 mW in 22nm-class cryo-CMOS (1\% of the 1.5W 4K budget), while Level 5 requires $\sim$2.5M gate-equivalents, the entire excess being runtime routing/allocation hardware; its route-resolution time exceeds the logical cycle beyond $\sim$100 logical qubits.

\textbf{FPGA Analytical Resource Model:} For Level 4 resource modeling, we use $LUTs = L_{decode} + L_{schedule} + L_{patch} + L_{meas}$,
where $L_{decode}$ is the number of LUTs required for decoding a logical instruction, $L_{schedule}$ for implementing a static or dynamic scheduler, $L_{patch}$ for tracking logical patch IDs with occupancy, and $L_{meas}$ for measurement status.

FPGA BRAM (Block RAM) size depends on the total number of instructions kept in memory, the maximum number of microcode operations among all expansions of all implemented instructions, and the bit width of each microcode operation: $BRAM = N_{instr} \times L_{max\ exp} \times W_{\mu op}$,
where $N_{instr}$ is the total number of instructions stored in memory, $L_{max\ exp}$ is the length of the longest micro-op sequence among available instructions, and $W_{\mu op}$ is the width, in bits, of a micro-op. The block nature of FPGA RAM forces us to size for the worst-case microcode expansion, yielding a larger resource requirement than a non-block SRAM/DRAM.

\section{Experimental Setup}
\label{sec:experimental_setup}
\subsection{Platform-Specific Parameters}
\label{sec:platform_params}
The bandwidth and power equations from Section~\ref{sec:bandwidth_power_model} are platform-agnostic, but the physical parameters ($BW_{cap}$, $P_{budget}$, $E_{bit}$), cycle times ($t_r$, $t_c$), and instruction widths vary across qubit platforms. We determine instruction and operand widths as the $\log_2$ of the number of operations and qubit counts, respectively, at each level.

\textbf{Superconducting Qubits (SC):} We model QEC on SC qubits using surface codes. Physical syndrome extraction rounds take $t_r = t_{init} + 4*t_{cnot} + t_{meas}$, where $t_{init}$, $t_{cnot}$, and $t_{meas}$ are the initialization, CNOT gate, and measurement times~\cite{fowler_gidney_2018}. A full logical operation requires $d$ rounds of error correction with logical cycle time $t_c = d \times t_r$. We use $t_{init} = 100ns$, $t_{cnot} = 50ns$, and $t_{meas} = 100ns$~\cite{resource_estimation_paper}.
SC architectures require the digital control engine (e.g., Cryo-CMOS) at the 4K stage of a dilution refrigerator, with the software driver at room temperature (300~K)~\cite{cryo_cmos_quantum}. The bounding parameter is the thermodynamic cooling capacity at 4~K, imposing \textbf{a strict power budget of 1.5W}~\cite{krinner}. State-of-the-art cryogenic transceivers crossing this thermal boundary operate at $E_{bit} \approx 0.5\text{-}5$~pJ/bit with \textbf{bandwidths of 50-60 Gbps}~\cite{9731563, 9141289, cqlink}; we use $E_{bit} =0.5$pJ/bit~\cite{cqlink}. $P_{budget}/E_{bit}$ upper-bounds interconnect bandwidth ($BW_{cap}$) at the 4K stage. We assume Aurora 8B/10B SerDes with encoding efficiency $\eta=0.8$~\cite{aurora}.

\textbf{Topological Majorana Qubits (TM):} TM architectures encode information in Majorana Zero Modes (MZMs) for hardware-level noise protection, built as tetrons using four MZMs per qubit~\cite{majorana_roadmap}. Physical operations are orchestrated via parity measurements; measurements take 100ns and return two bits per qubit. We use surface codes~\cite{Chao2020} and Hastings-Haah codes~\cite{Hastings2021, Haah2022} as QEC choices for TM. For surface code, the extraction round takes $t_r=20 \times t_{meas}$ with $t_c=d \times t_r$~\cite{Chao2020, Tran2020}; for Hastings-Haah code, $t_r=3 \times t_{meas}$ and $t_c=d \times t_r$~\cite{Paetznick2023}, but more physical qubits are used per logical qubit. Since physical operations are measurements, return bandwidth for Level 1 QISA is significantly higher than on SC platforms.
Like SC platforms, TM qubits reside at milli-Kelvin temperatures with control engines at the dilution fridge's 4K stage, subject to the same thermodynamic limit ($P_{budget} \approx 1.5$~W) and transceiver efficiency.


\subsection{Applications}
Our benchmarks are shown in Table~\ref{tab:benchmarks}: early quantum advantage experiments in condensed matter models, plus commercially relevant quantum chemistry and factoring applications using Q\# implementations from~\cite{qsharp}. For ease of benchmarking, larger applications such as factoring and quantum chemistry are split into sub-applications using representative kernels rather than a full algorithm: e.g., for modulo exponentiation, we sample a representative subset of chunks rather than applying windowed addition over all chunks of a 2048-bit register.
\begin{table}[t]
\centering
\caption{Benchmark applications and their problem sizes.}
\label{tab:benchmarks}
\begin{adjustbox}{width=\columnwidth,center}
\begin{tabular}{lrc}
\hline
\textbf{Application} & \textbf{Qubits} & \textbf{Cycles} \\
\hline
Ising Model                               & 100             & $3.6 \times 10^3$   \\
Hubbard Model                             & 100             & $9.9 \times 10^3$                  \\
Heisenberg Model                          & 100             & $1.1 \times 10^4$                  \\
Factoring Coset Initialization (2048 bit) & 9269            & $1.0 \times 10^5$                  \\
Modulo Exponentiation (2048 bit)          & 9851            & $1.9 \times 10^6$                  \\
QFT (2048 bit)                            & 2048            & $1.3 \times 10^5$                  \\ 
n2-10e-8e Quantum Chemistry (Walk operator)          & 257             & $2.4 \times 10^4$                     \\
polyyne-24e-24o Quantum Chemistry (Walk operator)    & 566             & $7.1 \times 10^5$                     \\
\hline
\end{tabular}
\end{adjustbox}
\end{table}


\subsection{Simulations}
We execute the benchmarks with two versions of the engine simulator. The ``realistic'' version, described in Section \ref{sec:modeling}, builds up layers of instructions in real time while executing one layer per cycle, even if more instructions could have been added in the future. The ``optimal'' version never flushes layers and instead builds them up indefinitely, ensuring maximal parallelism -- equivalent to a pre-compiled algorithm run with a perfect driver that decodes measurement results, tracks errors/corrections, and issues instructions in time without missing any cycle cutoff.

We run the simulations for QISA Levels 5, 4, and 3, recording instruction counts per cycle and total cycles executed, then compute issue/return bandwidth and transceiver power using the equations in Section \ref{sec:framework}. Levels 1 and 2 have very predictable instruction patterns dominated by idling QEC instructions, so their bandwidth is computed analytically from QEC circuit structure and Level 3 instruction counts.


\section{Results}
\label{sec:bandwidth_power}

\begin{figure*}[t]
  \centering
  \begin{minipage}[t]{0.48\textwidth}
    \begin{adjustbox}{width=0.90\columnwidth,center}
    \includestandalone{diagrams/bandwidth_issue_plot}
    \end{adjustbox}
  \end{minipage}
  \hfill
  \begin{minipage}[t]{0.48\textwidth}
    \begin{adjustbox}{width=0.90\columnwidth,center}
    \includestandalone{diagrams/bandwidth_return_plot}
    \end{adjustbox}
  \end{minipage}
  \caption{Bandwidth requirements for surface code-based superconducting qubit platform.} 
  \label{fig:bandwidth}
\end{figure*}
\begin{figure}[t]
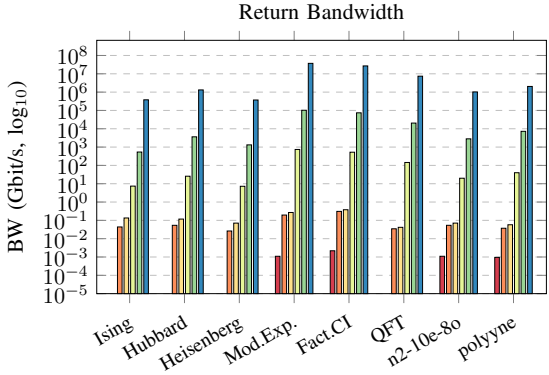

    \begin{adjustbox}{width=0.90\columnwidth,center}
    \includestandalone{diagrams/bandwidth_return_majorana_hh_plot}
    \end{adjustbox}
  \caption{Return bandwidth requirements for Hastings-Haah code based Majorana qubit platform.} 
  \label{fig:bandwidth_return_majorana_hh}
\end{figure}
\begin{figure}[t]
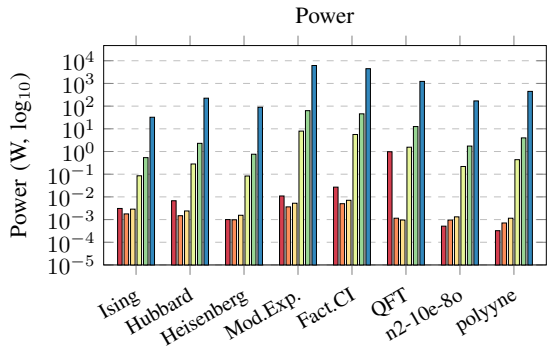

    \begin{adjustbox}{width=0.90\columnwidth,center}
    \includestandalone{diagrams/power_plot}
    \end{adjustbox}
    \caption{Maximum transceiver power for superconducting qubit platform.}
    \label{fig:power}
\end{figure}

\begin{table}[t]
  \caption{Simulated number of cycles for different benchmarks at different QISA levels.}
  \centering
  \footnotesize 
  \begin{tabular}{llrr}
    \toprule
    Benchmark & QISA level & Optimal Cycles & Realistic Cycles \\
    \midrule 
    Ising & 5 & 3,566& 10,007 \\
          & 4 & 5,443& 193,283 \\
          & 3 & 5,677& 208,426 \\
    Hubbard & 5 & 9,874& 205,379 \\
          & 4 & 156,245& 1,165,547 \\
          & 3 & 166,449& 1,298,602 \\
    Heisenberg & 5 & 11,072& 35,151 \\
          & 4 & 15,115& 499,174 \\
          & 3 & 15,667& 539,469 \\
    Mod.Exp. & 5 & 1,871,923&  2,178,688\\
          & 4 & 4,819,304&  27,061,488\\
          & 3 & 6,425,983& 34,852,439 \\
    Fact.C.I. & 5 & 102,376&  233,425\\
          & 4 & 183,076&  1,868,709\\
          & 3 & 280,724&  2,329,976\\
    QFT & 5 & 229,210& 117,244,631 \\
          & 4 & 60,541,211& 207,482,038 \\
          & 3 & 62,689,360& 223,750,284 \\
    n2 & 5 & 24,286&  26,871\\
          & 4 & 48,195&  73,138\\
          & 3 & 69,977&  103,106\\
    polyyne & 5 & 713,850&  783,954\\
          & 4 & 1,615,856&  2,632,952\\
          & 3 & 2,245,168&  3,716,700\\
    \bottomrule
  \end{tabular}
  \label{tab:layers}
\end{table}

\subsection{Level 1 Analysis}
Figure \ref{fig:bandwidth} shows issue bandwidth requirements at every QISA level. For Level 1 QISA, driver-engine bandwidth is massive: 55 Tbps to 20 Pbps across applications when addressing physical qubits individually, reducing to 700 Gbps-62 Tbps with SIMD instructions. This scales directly with total qubit count, since the driver issues physical QEC operations on every physical qubit every cycle. Even for the smallest scientifically interesting applications, such as the 100-logical-qubit Ising model, requirements exceed even room-temperature Infiniband (200 Gbps~\cite{infiniband}) and are orders of magnitude beyond the 50-60 Gbps limits of cryogenic transceivers~\cite{cqlink}.

Figure \ref{fig:bandwidth} also shows return bandwidth requirements, which are lower than issue bandwidth without SIMD (syndrome extraction uses more non-measurement instructions, and measurement outcomes need fewer bits than a typical instruction), but with SIMD control fanout range from 370 Gbps to 64 Tbps, similar to issue bandwidth.
For Majorana platforms, all Level 1 instructions are measurements, elevating return bandwidth for Level 1 and 2 QISAs; Figure \ref{fig:bandwidth_return_majorana_hh} shows Level 1 SIMD return bandwidth for Hastings-Haah code Majorana platforms ranging 533 Gbps-102 Tbps.



Figure \ref{fig:power} shows transceiver power draw estimates for different QISA levels. Similar to bandwidth, Level 1 causes the highest power demands: even with SIMD instructions, power draw violates thermal limits imposed by cryogenic control (1.5W), so while this level suits current experiments with hundreds of physical qubits, it is unsuited for scalable FTQC systems. SIMD operations are nonetheless very beneficial at the physical level, offering up to two orders of magnitude improvement in bandwidth and transceiver power.

\subsection{Level 2 Analysis}
Going from Level 1 SIMD to Level 2 QISA yields an average $6\times$ improvement in issue bandwidth, $12\times$ in return bandwidth, and $8\times$ in transceiver power, by abstracting the repeating syndrome-extraction pattern into one instruction. Level 2 still requires the driver to issue QEC instructions to idling qubits and process error syndromes, so bandwidth remains infeasible and transceiver power still violates cryogenic limits for large applications -- QISA assumptions in prior works such as XQSim~\cite{xqsim} are thus unrealizable in scalable systems.

\subsection{Level 3 Analysis}
Level 3 shifts QEC responsibility to the engine through implicit idling, freeing the driver from issuing instructions to idling qubits. Since idling qubits far outnumber active qubits in most cycles, this yields an average $214\times$ improvement in issue bandwidth, $490\times$ in return bandwidth, and $263\times$ in transceiver power over Level 2. Prior work introducing implicit idling~\cite{tannu_micro} evaluates issue bandwidths for defect-based surface codes on superconducting qubits, a suboptimal choice compared to lattice surgery; our work shows implicit idling is valuable for lattice surgery architectures too, across qubit technologies. From this level upward, transceiver power draw is within cooling limits for all benchmarks considered.

\subsection{Level 4 Analysis}
Abstracting further, from Level 3 to Level 4 provides average $1.5\times$ issue bandwidth, $2.6\times$ return bandwidth, and $1.6\times$ transceiver power improvement (up to $5.2\times$ return bandwidth in the best case). These gains arise because Level 3 instructions are frequently measurements, requiring the driver to maintain extra futures and issue correction gates; abstracting these away lets Level 4 ISAs cut return bandwidth without overloading the engine, which still needs only local, deterministically-timed operations.

Table \ref{tab:layers} reports cycle counts across QISA levels. Realistic cycles (program depth from driver-engine simulation) proxy application runtime and are typically scaled by cycle time in resource estimation studies~\cite{resource_estimation_paper}. Cycle counts at Level 4 are lower than at Level 3 across applications: for factoring kernels and quantum chemistry, we obtain $1.3$-$1.4\times$ speedup (on average $20\%$), important for long-running applications.

Beyond these measured gains, Level 4's abstraction over the QEC scheme is itself an architectural advantage: a single engine and instruction encoding serves surface-code and Hastings-Haah-code platforms alike and remains valid as new QEC codes are proposed, without re-exposing engine internals or re-verifying the driver each time.


\subsection{Level 5 Analysis}

At the extreme, abstracting from Level 4 to Level 5 QISA surprisingly worsens issue bandwidth by approximately $2\times$ and transceiver power by approximately $1.7\times$ despite issuing fewer instructions, for two reasons. First, Level 5 houses remote instructions that may act on any number of qubits (e.g., MeasPauli) and require much larger, operand-dependent instruction widths. Second, from Table \ref{tab:layers}, cycle counts drop sharply at this level because the instruction set assumes all-to-all logical qubit connectivity, and that remote CNOTs happen concurrently with the engine resolving path conflicts for Bell-pair based orchestration~\cite{edge_disjoint_paths}. Supporting this would require the engine to handle remote operation decomposition -- path finding, qubit mapping, and scheduling -- in hardware, which is better handled in software. Overall, Level 5 QISA are assumed in resource estimation studies \cite{littinski_game, resource_estimation_paper}; however, they are not practically feasible to implement in hardware.

Table \ref{tab:layers} compares optimal cycles (perfect driver) with realistic cycles. Our cycle-level simulation shows a significant average $7.3\times$ cycle count increase over optimal. Resource estimation studies \cite{littinski_game, resource_estimation_paper} scale optimal cycles by logical cycle time to predict runtime, and our work shows this significantly underestimates it. While compilation or parallelism-extraction strategies can improve cycle counts, our results underscore the need to move from analytical resource estimation to cycle-level simulation, as is standard in classical computer architecture.






\textbf{Overall, we recommend Level 4 QISAs. This level provides significant bandwidth and power reductions compared to lower-level ISAs and a clean abstraction over native QEC operations. It also offers operations with deterministic timing and fixed instruction widths, easing implementation on superconducting or Majorana platforms.}

\section{Related Work}
\label{sec:related} 
Early QISAs such as eQASM, cQASM, and OpenQASM~\cite{eqasm, cqasm, openqasm}, designed for the NISQ era, along with hardware-aware NISQ ISA recommendations~\cite{2312.05652, murali19, murali21}, inherently expose precise physical-level control to software. To model fault-tolerant control, simulators like XQSim~\cite{xqsim} mix abstraction levels, including both QEC-specific Level 2 micro-operations and QEC-agnostic Level 5 operations such as multi-qubit Pauli rotations; while these partially hide the low-level complexities of surface codes, they still impose significant bandwidth overheads. \cite{tannu_micro} proposes abstracting QEC issue operations away from software to reduce bandwidth (Level 3 in our framework) in older defect-based surface code architectures; we go beyond this by showing the importance of return bandwidth and further reducing software costs through Level 4 ISAs. Similarly, \cite{edge_disjoint_paths} uses a Level 3 ISA specific to surface codes, resulting in higher issue/return bandwidth requirements than our recommendation.

\section{Conclusions}
Practical FTQC depends not only on qubit quality or QEC theory, but equally on architecting the classical control interface that orchestrates them. Our work reveals a fundamental tension: the instruction set must be abstract enough to respect cryogenic thermal limits, yet not so abstract that it conceals logical-qubit management and inflates engine complexity. Level 4 Local Logical Operations resolves this tension, delivering physically realizable bandwidths and improved runtimes through a clean, deterministic engine interface, and motivates cycle-accurate simulation as standard FTQC design practice. Our recommendations generalize to any 2D connectivity QEC architecture on superconducting or topological Majorana qubits.

\section{Acknowledgements}
Claude Sonnet 4.6 was used for stylistic edits, drafting and figure preparation.

\bibliographystyle{IEEEtranS}
\bibliography{reference}

\end{document}

%% file: images/cnot_heir.tex
\begin{tikzpicture}[
    font=\sffamily,
    x=1cm, y=1cm,
    block/.style={draw=darkSlate, thick, rounded corners=3pt, fill=modernSlate, align=center, minimum height=0.7cm, minimum width=1.8cm, inner sep=2pt, font=\scriptsize},
    highlight/.style={draw=techIndigo, fill=modernIndigo!80, line width=1.2pt},
    levelbox/.style={draw=darkSlate!40, dashed, thick, rounded corners=4pt, inner sep=6pt},
    zoomline/.style={densely dashed, techIndigo!80, thick},
    marrow/.style={modernCoral, densely dashed, thick, ->, >=Stealth},
    cnot target/.style={circle, draw=darkSlate, thick, inner sep=0pt, minimum size=10pt, fill=modernSlate, path picture={\draw[thick, darkSlate] (path picture bounding box.north) -- (path picture bounding box.south) (path picture bounding box.east) -- (path picture bounding box.west);}}
]

\pgfdeclarelayer{background}
\pgfsetlayers{background,main}

\node[inner sep=3pt, anchor=north east] at (8.5, 1.3) {
    \tikz[baseline=-0.5ex]{\draw[marrow] (0,0) -- (0.5,0);} \tiny Measurement Return
};

\begin{scope}[local bounding box=L5]
    \path (-0.15, 0.4) rectangle (8.65, -0.4);
    \node[block, highlight, minimum width=3.5cm, font=\small] (l5b1) at (4.25, 0) {\textbf{Remote CNOT} ($Q_c \to Q_t$)};
\end{scope}

\node[levelbox, fit=(L5)] (box5) {};
\node[above=2pt of box5.north west, anchor=south west, font=\bfseries\scriptsize, text=darkSlate] {L5: Canonical Operations};

\begin{scope}[local bounding box=L4, yshift=-2cm]
    \path (-0.15, 0.4) rectangle (8.65, -2.6);
    
    \node[draw=darkSlate, fill=modernSlate, thick, rounded corners=3pt, minimum width=7.6cm, minimum height=0.6cm, font=\scriptsize\bfseries, align=center] (driver) at (4.25, 0) {Classical Driver: Process Parities \& Compute Pauli Updates};
    
    \node[block] (f1) at (0.8, -2.0) {Loop $O(N)$:\\Prep Pairs};
    \node[block] (f2) at (2.95, -2.0) {Loop $O(N)$:\\Bell Meas.};
    \node[block, highlight] (f3) at (5.35, -2.0) {\textbf{Local CNOTs}\\$Q_c{\to}A_1{\to}A_2$\\$A_5{\to}Q_t$};
    \node[block] (f4) at (7.55, -2.0) {\textbf{Measure}\\\textbf{Ends}};
    
    \draw[->, >=Stealth, thick, darkSlate] (f1) -- (f2);
    \draw[->, >=Stealth, thick, darkSlate] (f2) -- (f3);
    \draw[->, >=Stealth, thick, darkSlate] (f3) -- (f4);

    \draw[marrow] (f2.north) -- (f2.north |- driver.south) node[pos=0.4, right=-2pt, font=\tiny] {$m_{bsm}$};
    \draw[marrow] (f4.north) -- (f4.north |- driver.south) node[pos=0.4, right=-2pt, font=\tiny] {$m_{ends}$};

\end{scope}

\node[levelbox, fit=(L4)] (box4) {};
\node[above=2pt of box4.north west, anchor=south west, font=\bfseries\scriptsize, text=darkSlate] {L4: Local Logical (Teleportation Flow)};

\begin{scope}[local bounding box=L3, yshift=-6.8cm]
    \path (-0.15, 0.6) rectangle (8.65, -0.5);
    \node[block] (l3b1) at (1.06, 0) {Init Ancilla\\$A \to |+\rangle_L$};
    \node[block, highlight] (l3b2) at (3.18, 0) {Joint Parity\\$\mathcal{M}_{ZZ} (Q_c, A_1)$};
    \node[block] (l3b3) at (5.31, 0) {Joint Parity\\$\mathcal{M}_{XX} (A, A_1)$};
    \node[block] (l3b4) at (7.43, 0) {Measure $A$\\in $Z$-basis};
    
    \draw[->, >=Stealth, thick, darkSlate] (l3b1) -- (l3b2);
    \draw[->, >=Stealth, thick, darkSlate] (l3b2) -- (l3b3);
    \draw[->, >=Stealth, thick, darkSlate] (l3b3) -- (l3b4);
    
    \draw[marrow] (l3b2.north) -- ++(0, 0.4) node[above=-2pt, font=\tiny] {$m_{zz}$};
    \draw[marrow] (l3b3.north) -- ++(0, 0.4) node[above=-2pt, font=\tiny] {$m_{xx}$};
    \draw[marrow] (l3b4.north) -- ++(0, 0.4) node[above=-2pt, font=\tiny] {$m_a$};
\end{scope}

\node[levelbox, fit=(L3)] (box3) {};
\node[above=2pt of box3.north west, anchor=south west, font=\bfseries\scriptsize, text=darkSlate] {L3: QEC-Native Logical (Lattice Surgery)};

\begin{scope}[local bounding box=L2, yshift=-9.5cm]
    \path (-0.15, 0.6) rectangle (8.65, -0.5);
    \node[block, minimum width=2.4cm] (l2b1) at (1.5, 0) {Merge:\\Init seam $|0\rangle$};
    \node[block, highlight, minimum width=2.4cm] (l2b2) at (4.25, 0) {\textbf{Execute $\boldsymbol{d}$}\\ \textbf{QEC cycles}};
    \node[block, minimum width=2.4cm] (l2b3) at (7.0, 0) {Split:\\Measure seam $Z$};
    
    \draw[->, >=Stealth, thick, darkSlate] (l2b1) -- (l2b2);
    \draw[->, >=Stealth, thick, darkSlate] (l2b2) -- (l2b3);
    
    \draw[marrow] (l2b2.north) -- ++(0, 0.4) node[above=-2pt, font=\tiny] {$m_{synd}$};
    \draw[marrow] (l2b3.north) -- ++(0, 0.4) node[above=-2pt, font=\tiny] {$m_{seam}$};
\end{scope}

\node[levelbox, fit=(L2)] (box2) {};
\node[above=2pt of box2.north west, anchor=south west, font=\bfseries\scriptsize, text=darkSlate] {L2: QEC Micro-Operations (Rough Boundary Merge)};

\begin{scope}[local bounding box=L1, yshift=-11.4cm]
    \path (-0.15, 0.3) rectangle (8.65, -3.2);
    
    \foreach \i/\name in {1/vData 1, 2/vData 2, 3/vData 3, 4/vData 4} {
        \draw[thick, darkSlate] (1.2, -\i*0.5) node[left, font=\scriptsize] {\name} -- (8.0, -\i*0.5);
    }
    \draw[line width=1.0pt, darkSlate] (1.2, -2.8) node[left, font=\scriptsize] {vSynd S} -- (8.0, -2.8);
    
    \node[block, minimum height=0.5cm, minimum width=0.6cm, inner sep=2pt] (initz) at (2.2, -2.8) {$|0\rangle$};
    \node[above=1pt of initz, font=\bfseries\tiny, text=techIndigo] {vInitZ};
    
    \foreach \x/\y in {3.6/-0.5, 4.4/-1.0, 5.2/-1.5, 6.0/-2.0} {
        \filldraw[darkSlate] (\x, \y) circle (2pt);
        \draw[line width=1.0pt, darkSlate] (\x, \y) -- (\x, -2.8);
        \node[cnot target, minimum size=8pt] at (\x, -2.8) {};
    }
    \node at (4.8, -0.1) [font=\bfseries\tiny, text=techIndigo] {vCNOT array};
    
    \node[block, fill=modernCoral!20, minimum height=0.5cm, minimum width=0.6cm, inner sep=2pt] (measz) at (7.4, -2.8) {$\mathcal{M}_Z$};
    \node[below=1pt of measz, font=\bfseries\tiny, text=techIndigo] {vMeasZ};
    
    \draw[marrow] (measz.north) -- ++(0, 0.6) node[above=-2pt, font=\tiny] {$m_{z}$};
\end{scope}

\node[levelbox, fit=(L1)] (box1) {};
\node[above=2pt of box1.north west, anchor=south west, font=\bfseries\scriptsize, text=darkSlate] {L1: Physical (SIMD Weight-4 $Z$-Stabilizer Circuit)};

\begin{pgfonlayer}{background}
    \draw[zoomline] (l5b1.south west) -- (box4.north west);
    \draw[zoomline] (l5b1.south east) -- (box4.north east);
    
    \draw[zoomline] (f3.south west) -- (box3.north west);
    \draw[zoomline] (f3.south east) -- (box3.north east);

    \draw[zoomline] (l3b2.south west) -- (box2.north west);
    \draw[zoomline] (l3b2.south east) -- (box2.north east);
    
    \draw[zoomline] (l2b2.south west) -- (box1.north west);
    \draw[zoomline] (l2b2.south east) -- (box1.north east);
\end{pgfonlayer}

\end{tikzpicture}

%% file: images/tiled_surface_code.tex
\begin{tikzpicture}[
    font=\sffamily,
    x=1.3cm, y=1.3cm
]

    \node[font=\bfseries\small, text=darkSlate, anchor=north west] at (-0.6, 0.9) {Remote CNOT on $3 \times 6$ Logical Qubit Array};
    
    \foreach \col in {0,...,5} {
        \foreach \row in {0,...,2} {
            \node[draw=gray!30, fill=white, rounded corners=2pt, minimum size=1.1cm] at (\col, -\row) {};
        }
    }
    
    \node[draw=darkSlate, fill=techIndigo!70, thick, rounded corners=2pt, minimum size=1.1cm, text=white, font=\small\bfseries] (Qc) at (0, -1) {$Q_c$};
    \node[draw=darkSlate, fill=modernCoral!80, thick, rounded corners=2pt, minimum size=1.1cm, text=white, font=\small\bfseries] (Qt) at (5, -2) {$Q_t$};
    
    \foreach \col/\name in {1/A_1, 2/A_2, 3/A_3, 4/A_4, 5/A_5} {
        \node[draw=darkSlate, fill=modernIndigo!40, thick, rounded corners=2pt, minimum size=1.1cm, font=\small] (A\col) at (\col, -1) {$\name$};
    }

    \draw[techIndigo, line width=1.5pt, decorate, decoration={snake, amplitude=0.4mm, segment length=3mm}] ([yshift=6pt]A2.north) -- ([yshift=6pt]A3.north);
    \node[above=0pt, font=\scriptsize\bfseries, text=techIndigo] at (2.5, -0.4) {Bell Pair 1};
    
    \draw[techIndigo, line width=1.5pt, decorate, decoration={snake, amplitude=0.4mm, segment length=3mm}] ([yshift=6pt]A4.north) -- ([yshift=6pt]A5.north);
    \node[above=0pt, font=\scriptsize\bfseries, text=techIndigo] at (4.5, -0.4) {Bell Pair 2};
    
    \draw[modernCoral, line width=1.2pt, densely dashed, rounded corners=4pt] ([xshift=-3pt, yshift=-3pt]A3.south west) rectangle ([xshift=3pt, yshift=3pt]A4.north east);
    \node[below=2pt, font=\scriptsize\bfseries, text=modernCoral, align=center] at (3.5, -1.6) {Bell Meas};
    
    \draw[<->, thick, darkSlate] (Qc.north) to[bend left=30] node[above=1pt, font=\tiny\bfseries, align=center] {Local\\CNOT} (A1.north);
    \draw[<->, thick, darkSlate] (A1.south) to[bend right=30] node[below=1pt, font=\tiny\bfseries, align=center] {Local\\CNOT} (A2.south);
    \draw[<->, thick, darkSlate] (A5.east) to[bend left=45] node[right=1pt, font=\tiny\bfseries, align=center] {Local\\CNOT} (Qt.east);

\end{tikzpicture}

%% file: images/tool_diagram.tex
\begin{tikzpicture}[
    >=Stealth,
    boxbase/.style={draw=darkSlate, line width=1pt, rounded corners=4pt, align=center, font=\sffamily\small, inner sep=5pt},
    inputbox/.style={boxbase, fill=modernSlate, minimum width=3.2cm},
    stagebox/.style={boxbase, fill=modernSlate, minimum width=7cm},
    enginebox/.style={boxbase, fill=modernIndigo!20, minimum width=7cm},
    outputbox/.style={boxbase, fill=modernIndigo!20, minimum width=7cm},
    arrowlabel/.style={font=\sffamily\scriptsize, fill=white, inner sep=2pt, rounded corners=2pt},
    env/.style={rounded corners=6pt, inner sep=10pt, dashed, line width=0.6pt}
]

\node[inputbox, xshift=-2cm] (app_input) {
    \textbf{Application}\\[-1pt]
    \footnotesize Q\# program
};

\node[inputbox, xshift=2cm] (arch_config) {
    \textbf{Architecture Config}\\[-1pt]
    \footnotesize Qubit type, gate speeds,\\[-1pt]
    \footnotesize QISA level
};

\coordinate (input_center) at ($(app_input.south)!0.5!(arch_config.south)$);

\node[stagebox, below=1cm of input_center] (compiler) {
    \textbf{Q\# Compiler}\\[-1pt]
    \footnotesize Application $\rightarrow$ Canonical IR (Level 5)
};

\node[stagebox, below=0.3cm of compiler] (driver) {
    \textbf{ARIES Driver}\\[2pt]
    \footnotesize
    \begin{tabular}{l}
    $\bullet$ ARIES hierarchy transformations\\
    $\bullet$ Async instruction issue\\
    $\bullet$ Measurement processing\\
    $\bullet$ Pauli frame tracking
    \end{tabular}
};

\node[enginebox, below=0.3cm of driver] (engine) {
    \textbf{Engine Simulator}\\[2pt]
    \footnotesize
    \begin{tabular}{l}
    $\bullet$ Platform-aware execution timing\\
    $\bullet$ Bandwidth and power models
    \end{tabular}
};

\node[outputbox, below=0.6cm of engine] (outputs) {
    \textbf{Output Metrics}\\[2pt]
    \footnotesize
    \begin{tabular}{l}
    $\bullet$ Number of compute cycles\\
    $\bullet$ Bandwidth ($BW_{issue}$, $BW_{return}$)\\
    $\bullet$ SerDes power ($P_{TRX}$)\\
    $\bullet$ Measurement processing time
    \end{tabular}
};

\begin{scope}[on background layer]
    \node[env, draw=darkSlate!30, fill=modernSlate!50,
          fit=(compiler)(driver)(engine),
          inner xsep=10pt, inner ysep=8pt,
          label={[font=\sffamily\footnotesize\bfseries, text=darkSlate]above:Our Toolflow}] (tool_boundary) {};
\end{scope}

\draw[->, thick, draw=darkSlate!60] (app_input.south) -- (app_input.south |- tool_boundary.north);
\draw[->, thick, draw=darkSlate!60] (arch_config.south) -- (arch_config.south |- tool_boundary.north);


\draw[->, thick, draw=darkSlate!60] (engine) -- (outputs);

\end{tikzpicture}

%% file: diagrams/bandwidth_issue_plot.tex
\definecolor{isa5col}{HTML}{D53E4F}
\definecolor{isa4col}{HTML}{FC8D59}
\definecolor{isa3col}{HTML}{FEE08B}
\definecolor{isa2col}{HTML}{E6F598}
\definecolor{isa1.5col}{HTML}{99D594}
\definecolor{isa1col}{HTML}{3288BD}
  \begin{tikzpicture}
  \begin{axis}[
    title=Issue Bandwidth,
    width=\linewidth,
    height=5cm,
    ybar,
    bar width=0.084,
    enlarge x limits={abs=0.6},
    ylabel={BW (Gbit/s, log$_{10}$)},
    yticklabel={$10^{\pgfmathparse{\tick-2}\pgfmathprintnumber{\pgfmathresult}}$},
    ymin=0,
    ytick distance=1,
    xtick={0,1,2,3,4,5,6,7},
    xticklabels={Ising,Hubbard,Heisenberg,Mod.Exp.,Fact.CI,QFT,n2,polyyne},
    xticklabel style={rotate=30,anchor=north east,font=\small},
    legend style={at={(0.5,-0.32)},anchor=north,legend columns=6,font=\small},
    ymajorgrids=true,
    grid style=dashed,
    clip=false,
  ]
  \addplot+[fill=isa5col,draw=black,bar shift=-0.300]
    coordinates {(0,2.79) (1,3.13) (2,2.30) (3,3.34) (4,3.73) (5,5.29) (6,2.01) (7,1.81)};
  \addlegendentry{ISA-5}
  \addplot+[fill=isa4col,draw=black,bar shift=-0.180]
    coordinates {(0,2.51) (1,2.41) (2,2.25) (3,2.77) (4,2.89) (5,2.31) (6,2.18) (7,2.06)};
  \addlegendentry{ISA-4}
  \addplot+[fill=isa3col,draw=black,bar shift=-0.060]
    coordinates {(0,2.68) (1,2.60) (2,2.41) (3,2.94) (4,3.06) (5,2.20) (6,2.32) (7,2.27)};
  \addlegendentry{ISA-3}
  \addplot+[fill=isa2col,draw=black,bar shift=0.060]
    coordinates {(0,4.06) (1,4.58) (2,4.06) (3,6.02) (4,5.87) (5,5.31) (6,4.46) (7,4.76)};
  \addlegendentry{ISA-2}
  \addplot+[fill=isa1.5col,draw=black,bar shift=0.180]
    coordinates {(0,4.84) (1,5.35) (2,4.83) (3,6.80) (4,6.65) (5,6.09) (6,5.24) (7,5.54)};
  \addlegendentry{ISA-1(SIMD)}
  \addplot+[fill=isa1col,draw=black,bar shift=0.300]
    coordinates {(0,6.75) (1,7.58) (2,7.18) (3,9.02) (4,8.88) (5,8.32) (6,7.46) (7,7.89)};
  \addlegendentry{ISA-1}
  \end{axis}
  \end{tikzpicture}

%% file: diagrams/bandwidth_return_plot.tex
\definecolor{isa5col}{HTML}{D53E4F}
\definecolor{isa4col}{HTML}{FC8D59}
\definecolor{isa3col}{HTML}{FEE08B}
\definecolor{isa2col}{HTML}{E6F598}
\definecolor{isa1.5col}{HTML}{99D594}
\definecolor{isa1col}{HTML}{3288BD}
  \begin{tikzpicture}
  \begin{axis}[
    title=Return Bandwidth,
    width=\linewidth,
    height=5cm,
    ybar,
    bar width=0.084,
    enlarge x limits={abs=0.6},
    ylabel={BW (Gbit/s, log$_{10}$)},
    yticklabel={$10^{\pgfmathparse{\tick-4}\pgfmathprintnumber{\pgfmathresult}}$},
    ymin=0,
    ytick distance=1,
    xtick={0,1,2,3,4,5,6,7},
    xticklabels={Ising,Hubbard,Heisenberg,Mod.Exp.,Fact.CI,QFT,n2,polyyne},
    legend style={at={(0.5,-0.32)},anchor=north,legend columns=6,font=\small},
    xticklabel style={rotate=30,anchor=north east,font=\small},
    ymajorgrids=true,
    grid style=dashed,
    clip=false,
  ]
  \addplot+[fill=isa5col,draw=black,bar shift=-0.300]
    coordinates {(0,nan) (1,nan) (2,nan) (3,1.90) (4,2.20) (5,nan) (6,1.90) (7,1.84)};
  \addlegendentry{ISA-5}
  \addplot+[fill=isa4col,draw=black,bar shift=-0.180]
    coordinates {(0,3.51) (1,3.59) (2,3.28) (3,4.15) (4,4.35) (5,3.40) (6,3.59) (7,3.43)};
  \addlegendentry{ISA-4}
  \addplot+[fill=isa3col,draw=black,bar shift=-0.060]
    coordinates {(0,3.99) (1,3.93) (2,3.71) (3,4.29) (4,4.44) (5,3.48) (6,3.71) (7,3.62)};
  \addlegendentry{ISA-3}
  \addplot+[fill=isa2col,draw=black,bar shift=0.060]
    coordinates {(0,5.74) (1,6.27) (2,5.73) (3,7.73) (4,7.59) (5,7.03) (6,6.16) (7,6.46)};
  \addlegendentry{ISA-2}
  \addplot+[fill=isa1.5col,draw=black,bar shift=0.180]
    coordinates {(0,6.57) (1,7.36) (2,6.92) (3,8.81) (4,8.67) (5,8.11) (6,7.25) (7,6.77)};
  \addlegendentry{ISA-1(SIMD)}
  \addplot+[fill=isa1col,draw=black,bar shift=0.300]
    coordinates {(0,7.94) (1,8.80) (2,8.38) (3,10.25) (4,10.11) (5,9.55) (6,8.69) (7,9.12)};
    \addlegendentry{ISA-1}
  \end{axis}
  \end{tikzpicture}

%% file: diagrams/bandwidth_return_majorana_hh_plot.tex
\definecolor{isa5col}{HTML}{D53E4F}
\definecolor{isa4col}{HTML}{FC8D59}
\definecolor{isa3col}{HTML}{FEE08B}
\definecolor{isa2col}{HTML}{E6F598}
\definecolor{isa1.5col}{HTML}{99D594}
\definecolor{isa1col}{HTML}{3288BD}
  \begin{tikzpicture}
  \begin{axis}[
    title=Return Bandwidth,
    width=\linewidth,
    height=5.6cm,
    ybar,
    bar width=0.084,
    enlarge x limits={abs=0.6},
    ylabel={BW (Gbit/s, log$_{10}$)},
    yticklabel={$10^{\pgfmathparse{\tick-5}\pgfmathprintnumber{\pgfmathresult}}$},
    ymin=0,
    ytick distance=1,
    xtick={0,1,2,3,4,5,6,7},
    xticklabels={Ising,Hubbard,Heisenberg,Mod.Exp.,Fact.CI,QFT,n2-10e-8o,polyyne},
    xticklabel style={rotate=30,anchor=north east,font=\small},
    legend style={at={(0.5,-0.32)},anchor=north,legend columns=6,font=\small},
    ymajorgrids=true,
    grid style=dashed,
    clip=false,
  ]
  \addplot+[fill=isa5col,draw=black,bar shift=-0.300]
    coordinates {(0,nan) (1,nan) (2,nan) (3,2.04) (4,2.34) (5,nan) (6,2.04) (7,1.98)};
  \addplot+[fill=isa4col,draw=black,bar shift=-0.180]
    coordinates {(0,3.64) (1,3.73) (2,3.42) (3,4.29) (4,4.49) (5,3.54) (6,3.73) (7,3.57)};
  \addplot+[fill=isa3col,draw=black,bar shift=-0.060]
    coordinates {(0,4.13) (1,4.07) (2,3.85) (3,4.43) (4,4.58) (5,3.62) (6,3.85) (7,3.76)};
  \addplot+[fill=isa2col,draw=black,bar shift=0.060]
    coordinates {(0,5.87) (1,6.41) (2,5.86) (3,7.87) (4,7.72) (5,7.16) (6,6.30) (7,6.60)};
  \addplot+[fill=isa1.5col,draw=black,bar shift=0.180]
    coordinates {(0,7.73) (1,8.56) (2,8.12) (3,10.01) (4,9.87) (5,9.31) (6,8.45) (7,8.86)};
  \addplot+[fill=isa1col,draw=black,bar shift=0.300]
    coordinates {(0,10.58) (1,11.12) (2,10.57) (3,12.57) (4,12.43) (5,11.87) (6,11.01) (7,11.31)};
  \end{axis}
  \end{tikzpicture}

%% file: diagrams/power_plot.tex
\definecolor{isa5col}{HTML}{D53E4F}
\definecolor{isa4col}{HTML}{FC8D59}
\definecolor{isa3col}{HTML}{FEE08B}
\definecolor{isa2col}{HTML}{E6F598}
\definecolor{isa1.5col}{HTML}{99D594}
\definecolor{isa1col}{HTML}{3288BD}

  \begin{tikzpicture}
  \begin{axis}[
    title={Power},
    width=0.95\linewidth,
    height=4.9cm,
    ybar,
    bar width=0.084,
    enlarge x limits={abs=0.6},
    ylabel={Power (W, log$_{10}$)},
    yticklabel={$10^{\pgfmathprintnumber{\pgfmathresult}}$},
    ytick distance=1,
    ytick scale label code/.code={},
    yticklabel style={/pgf/number format/.cd,fixed,precision=1},
    yticklabel={$10^{\pgfmathparse{\tick-5}\pgfmathprintnumber{\pgfmathresult}}$},
    ymin=0,
    xtick={0,1,2,3,4,5,6,7},
    xticklabels={Ising,Hubbard,Heisenberg,Mod.Exp.,Fact.CI,QFT,n2-10e-8o,polyyne},
    xticklabel style={rotate=30,anchor=north east,font=\small},
    legend style={at={(0.5,-0.32)},anchor=north,legend columns=6,font=\small},
    ymajorgrids=true,
    grid style=dashed,
    clip=false,
  ]
  \addplot+[fill=isa5col,draw=black,bar shift=-0.300]
    coordinates {(0,2.49) (1,2.83) (2,2.00) (3,3.04) (4,3.43) (5,4.99) (6,1.71) (7,1.51)};
  \addplot+[fill=isa4col,draw=black,bar shift=-0.180]
    coordinates {(0,2.25) (1,2.17) (2,1.99) (3,2.56) (4,2.70) (5,2.06) (6,1.98) (7,1.85)};
  \addplot+[fill=isa3col,draw=black,bar shift=-0.060]
    coordinates {(0,2.46) (1,2.38) (2,2.19) (3,2.72) (4,2.85) (5,1.98) (6,2.12) (7,2.06)};
  \addplot+[fill=isa2col,draw=black,bar shift=0.060]
    coordinates {(0,3.93) (1,4.45) (2,3.92) (3,5.90) (4,5.75) (5,5.19) (6,4.34) (7,4.64)};
  \addplot+[fill=isa1.5col,draw=black,bar shift=0.180]
    coordinates {(0,4.73) (1,5.36) (2,4.88) (3,6.80) (4,6.66) (5,6.10) (6,5.24) (7,5.60)};
  \addplot+[fill=isa1col,draw=black,bar shift=0.300]
    coordinates {(0,6.51) (1,7.35) (2,6.95) (3,8.79) (4,8.65) (5,8.09) (6,7.23) (7,7.65)};
  \end{axis}
  \end{tikzpicture}